# Quantitative evaluation of sense of discrepancy to operation response using event-related potential

Kazutaka UEDA*, Yuki SAKAI* and Hideyoshi YANAGISAWA*
*Department of Mechanical Engineering,
Graduate School of Engineering, The University of Tokyo
7-3-1 Hongo, Bunkyo-ku, Tokyo, 113-8656 Japan
E-mail: ueda@design-i.t.u-tokyo.ac.jp

**Abstract**

This study aimed to develop a method to evaluate the sense of discrepancy to the operation response quantitatively. We examined the availability of event-related potential (P300), which is considered to reflect attention to stimulation, to evaluate the sense of discrepancy to the product response to the user's action. In the experiment using subjective evaluation and P300 to investigate the sense of discrepancy due to the lack of operation response (sound and vibration) to the shutter operation of the mirrorless single-lens camera, it was confirmed that P300 amplitude corresponds to the degree of the subjective sense of discrepancy. Our results showed that the P300 amplitude could evaluate the sense of discrepancy to the operation response.

## 1. Introduction

In many conventional products, the user has perceived sensory information such as sounds and vibrations generated by the operation of mechanical mechanisms as the operation response to the operation input. In recent years, due to the simplification and digitization of products, when mechanical operation responses are missing, the user cannot obtain sensory information associated with the operation responses, and problems such as failure to obtain confirmation for operations occur. Taking a camera as an example, in a conventional single-lens reflex camera, when the shutter is released, sound and vibration are emitted when the shutter curtain descends. On the other hand, due to digitization, mirrorless single-lens cameras do not generate sound and vibration. In order to solve the problem, it is conceivable to add digital sound and vibration to the product, but it is necessary to design an operation response that does not give the user a sense of discrepancy. In this study, we aimed to construct a method to quantitatively evaluate the user's sense of discrepancy with product operation response. We examined the availability of event-related potential (P300), which is considered to reflect attention to stimuli [1-5], as a quantitative evaluation method of sense of discrepancy. P300 is a positive component of the electroencephalogram (EEG) that appears predominantly at the parietal area with a latency of approximately 250-600 ms after stimulus presentation. We considered that the P300 amplitude would increase due to the sense of discrepancy with the stimulus. We used a mirrorless single-lens camera to investigate the sense of discrepancy caused by the lack of operation response (vibration, sound) to the release button operation using a four-level Likert scale and measurements of P300 amplitudes.

## 2. Experimental methods

### 2.1 Participants

Right-handed healthy volunteers (n = 7; 5 men and 2 women; 22 - 26 years) who neither had diseases nor brain-related disorders participated. All participants had no professional use of the camera. The study protocol was approved by the Ethics Committee of the Graduate School of Engineering, the University of Tokyo. All participants provided written informed consent prior to their participation in this study.

## 2.2 Apparatus and Stimuli

The experimental setup was configured as in Fig. 1 using a mirrorless camera (α7MarkⅡ, Sony, Tokyo, Japan). When the participant pressed the release button, vibration and sound were presented. Input by button press and the stimulus presentation were controlled using the stimulus presentation control software (Presentation, Neurobehavioral Systems, Inc., California, U.S.A.). Vibration stimulus (frequency 320 Hz, square wave with 120 ms duration) was amplified by an amplifier (Power Amplifier Type 2706, Brüel & Kjær, Nærum, Denmark) and presented from a linear actuator (9×10×22.6 mm) placed inside the camera grip. The shutter sound stimulus used the shutter sound (120 ms duration) used in α7 Mark II. A noise canceling earphone (QuietComfort20, Bose, Massachusetts, U.S.A.) was used to present the shutter sound. The participants wore earmuffs to eliminate the influence of sound caused by vibration. The shutter sound was presented at 74 dB.

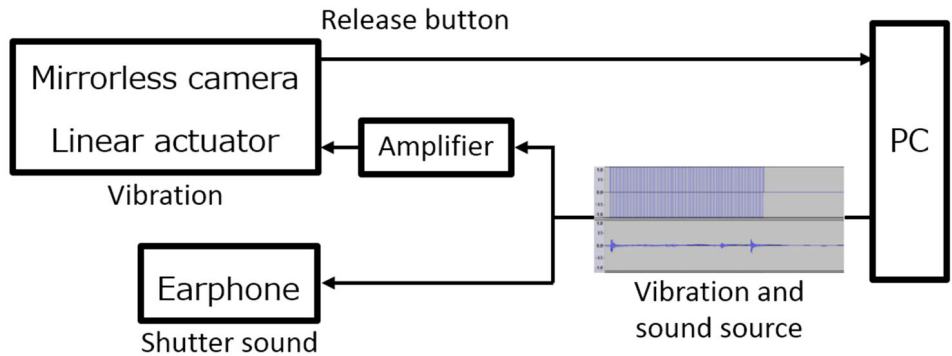

Fig. 1 Outline of experimental setup

## 2.3 Procedure

After participants received an explanation of the experiment, an operation task was conducted in the following order while participants sat in a chair in a comfortable position (Fig. 2). Participants pressed the release button at intervals of 1 to 2 seconds. A neutral landscape photograph that does not evoke emotion was presented on the display. Every time the release button is pressed, vibration stimulation and shutter sound stimulation are presented. There were 4 types of operation responses to the participant pressing the release button: trials in which both shutter sound and vibration stimulus were presented (audio and tactile, 70 trials) as standard response, and trials in which only shutter sound was presented (audio only, 10 trials), trials in which only vibration stimulus was presented (tactile only, 10 trials), and trials in which nothing was presented (none, 10 trials) as deviant responses. Four operation responses were presented in random order.

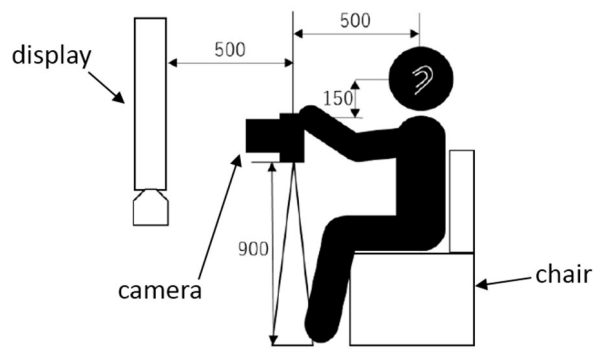

Fig. 2 Outline of experimental setup (unit: mm)

## 2.4 EEG recordings

The EEG data were recorded during tasks using a digital recorder (Polymate AP1132, TEAC Corporation, Tokyo,

Japan) and an EEG recording software (AP Monitor, TEAC Corporation, Tokyo, Japan). EEGs were recorded from 30 active electrodes on the scalp by the international 10-20 system (Fp1/Fp2, F7/F8, F3/F4, Fz, FT7/ FT8, FC3/ FC4, FCz, T7/ T8, C3/ C4, Cz,TP7/TP8, CP3/ CP4, CPz, P7/ P8, P3/ P4, Pz, O1/ O2, and Oz, Fig. 3) [6] and from the nose top as reference. The data were recorded at a sampling frequency of 500 Hz. The time constant was set at 3 s. All electrode impedances were below 50 kΩ.

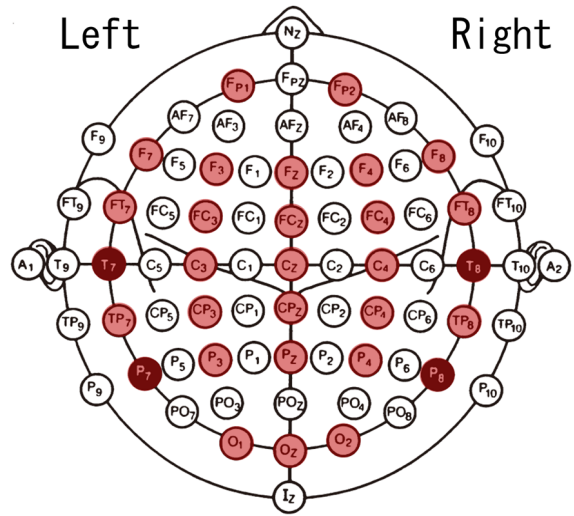

Fig. 3   EEG electrode placement (red circles). The diagram is a modification of [6].

**2.5 EEG data analysis**

The EEG data were exported to EEGLAB14.1.1b (MATLAB toolbox) [7] for ERP analysis, and bandpass filter of 0.1–20 Hz was applied. The data were segmented into epochs spanning from the period starting 200 ms before the button pressing onset and ending 1000 ms after the button pressing onset. This epoching was done separately for each participants and each type of operation response. Extended independent component analysis (ICA) was conducted using the runica algorithm [8] implemented in EEGLAB. The resultant EEG artifacts accounting for eye movement and muscle potentials were removed from the data. The ERP waveforms were obtained by averaging data from the period starting 200 ms before the button pressing onset and ending 1000 ms after the button stimulus onset. For each averaged waveform, the 200-ms period preceding the stimulus onset was defined as the baseline. The P300 component was designated as the largest positive peak occurring 250–600 ms after the button pressing. The baseline-to-peak P300 amplitudes were measured at the Pz point, which was the dominant electrode site. The P300 amplitude for Pz electrodes with regard to each type of operation response was used as the dependent variable for repeated measures analysis of variance (ANOVA).

**3. Results and Discussion**

Fig. 4 shows the average Likert scale score for the sense of discrepancy for each type of operation response. The error bars on the score represent the standard error. Compared with audio and tactile, and audio only, tactile only and none were higher scores. It was shown that the sense of discrepancy was subjectively high when there was no sound as the camera operation response. A one-factor ANOVA was performed with the type of operation response as an explanatory variable and the score for the sense of discrepancy as an objective variable. The type of operation response did not significantly affect the score for the sense of discrepancy. Fig. 5 shows the average P300 amplitude for each type of operation response. The error bars represent the standard error. Compared with audio and tactile, audio only, tactile only, and none were larger. A one-factor ANOVA was performed with the type of operation response as an explanatory variable and P300 amplitudes as an objective variable. The main effect of the type of operation response was significant. As a result of the multiple comparison, the P300 amplitudes evoked by sound only and none were larger than that evoked by audio and tactile were shown. The subjective sense of discrepancy and the P300 amplitude with the camera operation response showed the same tendency. In previous research [5], it was reported that as a surprise to an event, attention turned to the event, and the P300 amplitude increased, and P300 can be used as an indicator of surprise. In the results of this study, when the operation response to the release button press of the camera is unnatural, attention is directed to the operation response, and it is thought that the P300 amplitude increases. From these results, it was possible to show that the P300 amplitude can evaluate the sense of discrepancy to the operation response.

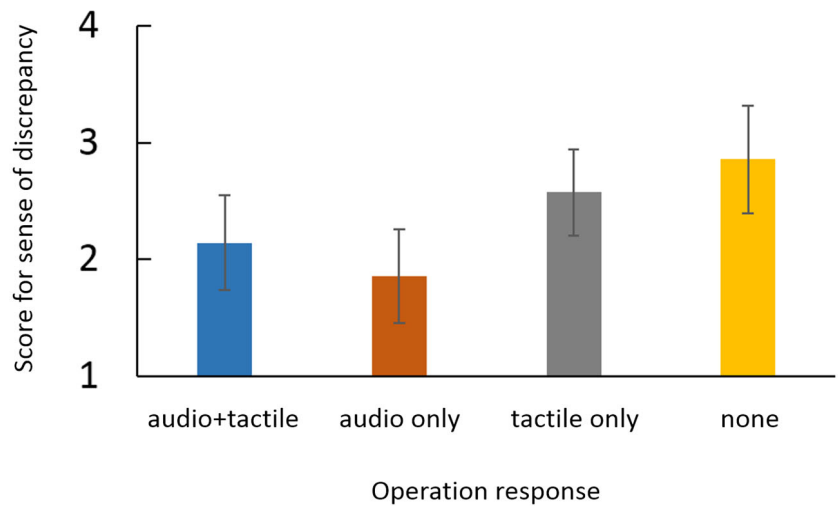

Fig. 4   Subjectively reported scores for sense of discrepancy (n=7).

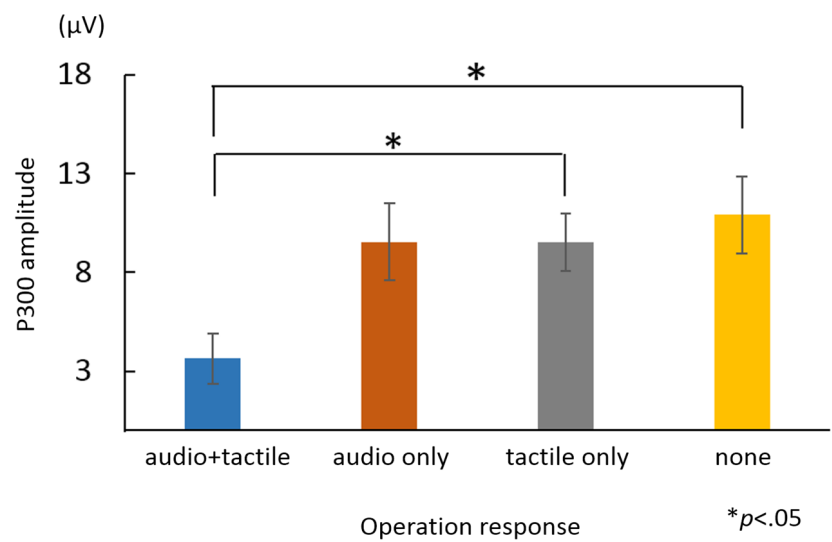

Fig. 5   P300 amplitudes for each type of operation response (n=7).